\documentclass[
preprint,
showpacs,amsmath,amssymb,superscriptaddress]{revtex4}

\usepackage{amsmath,amsfonts,amssymb}
\usepackage{bm} 
\usepackage[dvips]{graphicx}
\begin{document}
\title{Current-perpendicular-to-plane magnetoresistance of a domain wall 
confined in a nano-oxide-layer 
}
\author{Jun Sato}
\email{jun-sato@aist.go.jp}
\affiliation{Nanotechnology Research Institute, AIST, Tsukuba 305-8568, Japan}
\author{Katsuyoshi Matsushita}
\email{k-matsushita@aist.go.jp}
\affiliation{Nanotechnology Research Institute, AIST, Tsukuba 305-8568, Japan}
\author{Hiroshi Imamura}
\email{h-imamura@aist.go.jp}
\affiliation{Nanotechnology Research Institute, AIST, Tsukuba 305-8568, Japan}
%
\date{February 27, 1624}
\begin{abstract}
We theoretically study the current-perpendicular-to-plane magnetoresistance 
of a domain wall confined in a current-confined-path (CCP) structure 
made of a nano-oxide-layer (NOL). 
In order to calculate the MR ratio of the system, 
the continuity equations for charge and spin currents are numerically
solved with the three-dimensional CCP geometry by use of finite
element method.  It is confirmed that the MR ratio is enhanced by the 
CCP structure, which is consistent with the experimental results.
\end{abstract}
\maketitle
Current-perpendicular-to-plane giant
magnetoresistance (CPP-GMR) has attracted much attention for its
potential application as a read sensor for high-density magnetic recording. 
In order to realize a high-density magnetic
recording, we need MR devices with high MR ratio and low resistance
area product (RA). Although the RA value of a CPP-GMR system is much
smaller than that of a tunneling magnetoresistance (TMR) system, the
MR ratio of a conventional CPP-GMR system still remains a small value
of a few \%.  Much effort has been devoted to increasing the MR
ratio of the CPP-GMR system.
One of the candidates for such a device is 
the CPP-GMR spin-valve with current-confined-paths 
made of a nano-oxide-layer (NOL) with a lot of small
metallic channels \cite{fukuzawa04, fukuzawa05}. 
The MR ratio of 10.2\% is obtained by the CPP spin-valve with non-magnetic Cu nanocontacts. 
This MR enhancement by the CCP structure was 
theoretically studied in detail in Refs. \cite{imamura07, smi08}. 

Recently, Fuke {\it et al.} showed that 
such MR enhancement appears even in the system with ferromagnetic nanocontacts. 
They obtained the MR ratio of 7-10\% 
by the CoFe nanocontacts in NOL with CCP structure \cite{fuke07}. 
It seems that this MR effect originates from 
the spin accumulation around the domain wall confined in the NOL. 

In order to analyze this experimental result, 
we calculate the MR ratio of the CCP system with a domain wall.  
The MR ratio is determined from the distribution of the spin accumulation 
around the domain wall, 
which is obtained by solving the spin diffusion equation \cite{valet93, zlf} 
with the use of finite element method \cite{ram-mohan,ichimura07}. 

Let us begin with a brief introduction to the theory 
of the spin diffusion equation \cite{zlf}, 
which we solve to determine the MR ratio of the system. 
According to the Ref. \cite{zlf}, 
the electric current $\vec{j_e}$ and the spin current $\vec{\mathbf{j}}_m$ 
can be written as 
\begin{align}
\label{current1}
&\vec{j}_e=2C_0\vec{E}-2\mathbf{D}\cdot\vec{\nabla}\mathbf{m}, \\
\label{current2}
&\vec{\mathbf{j}}_m=2\mathbf{C}\vec{E}-2D_0\vec{\nabla}\mathbf{m}, 
\end{align}
where we choose the unit such as the charge of an electron and the Bohr magneton 
are set to be 1: $e=\mu_B=1$. 
Here we omit the term proportional to the derivative of the charge accumulation. 
The vectors denoted by the arrow $\vec{v}$ represent those in real space 
and the bold font $\mathbf{v}$ represent those in the spin space. 
$\vec{E}$ is the electric field and $\mathbf{m}$ is the spin accumulation. 
The conductivity $\hat{C}$ and the diffusion constant $\hat{D}$ 
are written in the spinor form as $\hat{C}=C_0+\mathbf{\sigma}\cdot\mathbf{C}$, 
$\hat{D}=D_0+\mathbf{\sigma}\cdot\mathbf{D}$, 
where the ${\bf \sigma}$ are the Pauli matrices. 
$\mathbf{C}$ and $\mathbf{D}$ are proportional to the spin polarization parameters 
$\beta$ and $\beta'$ as 
$\mathbf{C}=\beta C_0 \mathbf{M}$ and $\mathbf{D}=\beta' D_0 \mathbf{M}$, 
where $\mathbf{M}$ is the unit vector in the direction of the local magnetization. 

In order to derive the equation to determine the spin accumulation $\mathbf{m}$, 
we eliminate the electric field $\vec{E}$ from the equations 
(\ref{current1}) and (\ref{current2}) and obtain
\begin{align}
\vec{\mathbf{j}}_m=\beta\,\mathbf{M}\,\vec{j}_e-2D_0\left[
\vec{\nabla}\mathbf{m}-\beta\beta'\mathbf{M}
\left(\mathbf{M}\cdot\vec{\nabla}\mathbf{m}\right)
\right]. 
\end{align}
For convenience we introduce an operator $\hat{\mathbf{A}}$ acting on the spin space as 
\begin{align}
\hat{\mathbf{A}}=-2D_0\left[\!
\begin{pmatrix}1&&\\&1&\\&&1\end{pmatrix}
\!-\!\beta\beta'\!\begin{pmatrix}
M_1^2&M_1M_2&M_1M_3\\M_2M_1&M_2^2&M_2M_3\\M_3M_1&M_3M_2&M_3^2
\end{pmatrix}\!\right], 
\end{align}
where we write the components of the local magnetization as 
$\mathbf{M}=(M_1, M_2, M_3)$. 
With the use of this matrix $\hat{\mathbf{A}}$, 
the expression of the spin current is simplified into 
\begin{align}
\vec{\mathbf{j}}_m=\beta\,\mathbf{M}\,\vec{j}_e
+\hat{\mathbf{A}}\vec{\nabla}\mathbf{m}.
\end{align}

The continuity equation for the spin accumulation 
takes the form \cite{zlf}
\begin{align}
\frac{\partial\mathbf{m}}{\partial t}
=-\vec{\nabla}\cdot\vec{\mathbf{j}}_m
-(J/\hbar)\mathbf{m}\times\mathbf{M}-\frac{\mathbf{m}}{\tau_{\text{sf}}}, 
\label{motion}
\end{align}
where $J$ is the coupling constant of the interaction between the spin accumulation and 
the local moment, 
and $\tau_{\text{sf}}$ is the spin-flip relaxation time of the conduction electron. 
In order to obtain the solution for the stationary state, 
we set the left hand side of Eq. (\ref{motion}) to zero, 
\begin{align}
0&=\vec{\nabla}\cdot\vec{\mathbf{j}}_m
+(J/\hbar)\mathbf{m}\times\mathbf{M}+\frac{\mathbf{m}}{\tau_{sf}} \nonumber\\
&=\vec{\nabla}\cdot\left(\beta\,\mathbf{M}\,\vec{j}_e
+\hat{\mathbf{A}}\vec{\nabla}\mathbf{m}\right)
+(J/\hbar)\mathbf{m}\times\mathbf{M}+\frac{\mathbf{m}}{\tau_{sf}}. 
\end{align}
Introducing the matrix $\hat{\mathbf{B}}$ on the spin space as 
\begin{align}
\hat{\mathbf{B}}=\frac1{\tau_{sf}}\begin{pmatrix}1&&\\&1&\\&&1\end{pmatrix}
+(J/\hbar)\begin{pmatrix}0&M_3&-M_2\\-M_3&0&M_1\\M_2&-M_1&0\end{pmatrix}, 
\end{align}
we obtain the equation for the spin accumulation $\mathbf{m}$ to be solved 
\begin{align}
\vec{\nabla}\cdot\left(\beta\,\mathbf{M}\,\vec{j}_e
+\hat{\mathbf{A}}\vec{\nabla}\mathbf{m}\right)
+\hat{\mathbf{B}}\mathbf{m}=0. 
\label{eq1}
\end{align}

The determination of the MR ratio of the system consists of the following three steps. 
First we determine the charge current density $\vec{j}_e$ from the 
continuity equation $\vec{\nabla}\cdot\vec{j}_e=0$. 
Next we solve the Eq. (\ref{eq1}) 
for a given local magnetization $\mathbf{M}$ in the domain wall 
and determine the spin accumulation $\mathbf{m}$. 
Finally we determine the voltage drop of the system by integrating the 
electric field $\vec{E}=(\vec{j}_e+2\mathbf{D}\cdot\vec{\nabla}\mathbf{m})/(2C_0)$. 
Since we apply the constant electric current density, 
the voltage drop is proportional to the total resistance, 
from which we obtain the MR ratio of the system.

\begin{figure}[p]
  \includegraphics[width=\columnwidth]{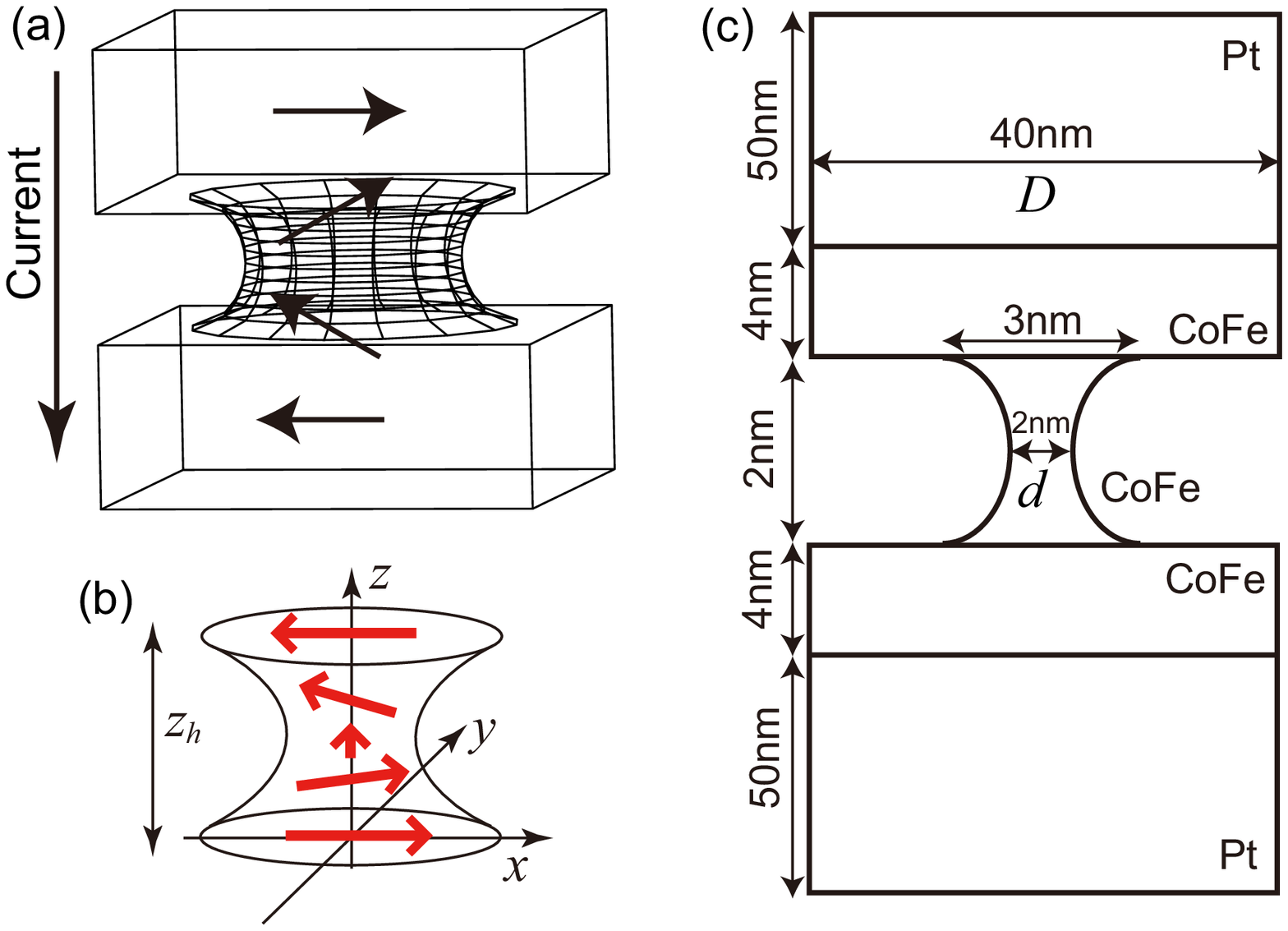}
  \caption{
  (a) A CCP spin-valve with a domain wall is schematically shown.
   A domain wall is confined in a nanocontact.  
  (b) The close-up of the contact region is schematically shown. 
   We adopt Bloch domain wall as a magnetic structure in the nanocontact
   $\mathbf{M}=(\cos(\pi z/z_h),\sin(\pi z/z_h),0)$. 
  (c) The cross-sectional view of the CCP spin-valve is
  schematically shown. The curve of the contact region is modeled by
  the half ellipse,  which is tangent to the ferromagnetic layers. 
   In our calculation we keep the shape of the ellipse fixed and vary the
   size of the contact by changing the contact diameter $d$.
   We also vary the size of the electrode $D$ with fixed contact diameter $d$. 
   }
  \label{fig1}
\end{figure}
Now we move onto the analysis of the MR ratio of the CCP spin-valve with a domain wall. 
The system we consider is schematically shown in Fig. \ref{fig1} (a).
A domain wall is confined in a nanocontact, 
which is sandwiched between pinned layer and free layer. 
We adopt Bloch domain wall as a magnetic structure in the nanocontact. 
The local magnetization $\mathbf{M}$ is given by
$\mathbf{M}=(\cos(\pi z/z_h),\sin(\pi z/z_h),0)$, 
where $z_h$ is the thickness of the nanocontact (=2nm) and 
$z$ is the height from the bottom of the nanocontact as is shown in Fig. \ref{fig1} (b). 

The cross-sectional view of the system is shown in Fig. \ref{fig1} (c). 
The curve of the contact region is modeled by the half ellipse, 
which is tangent to the ferromagnetic layer. 
The length of the major axis of the ellipse, 
which is the same as the thickness of the contact, is taken to be 2 nm. 
The length of the semi-minor axis of the ellipse is 1 nm.
The pinned and free layers are assumed to be 40nm$\times$40nm$\times$4nm rectangles. 
We also consider the Pt capping layer with the thickness 50nm. 
We set the value of the resistivity of Pt layer as large as $1\Omega\mu\text{m}$ 
so that the total RA without CCP structure is $0.1\Omega\mu\text{m}^2$ 
which is the value reported in the experiments \cite{fukuzawa04, fukuzawa05}. 

The spin accumulation $\mathbf{m}$ is determined by numerically solving Eq. 
\eqref{eq1} by use of the finite element method \cite{ram-mohan}, \cite{ichimura07}.
The system is divided into hexahedral elements and
the total number of elements is of the order of 10$^4$.
We assume that the free and pinned layers and the nanocontact are all made of CoFe. 
We use material parameters for the conventional CPP-GMR spin-valve system; 
 $\beta=0.65$, 
 $\rho_{\rm CoFe}=150\,\Omega\,\text{nm}$, 
 $\lambda_{\rm CoFe}=12\,\text{nm}$, 
 $\lambda_{\rm Pt}=14\,\text{nm}$, 
 $J=0.5\,\text{eV}$, 
where $\lambda$ denotes the spin diffusion length. 
Especially we adopt the value of resistivity of CoFe in the contact region as 
1$\Omega\mu$m, which is the value reported in the experiments \cite{fukuzawa04, fukuzawa05}. 
This enhancement of the resistivity in the contact region is due to the 
decrease of the purity of CoFe in the oxidation process. 
The diffusion constant $D_0$ and the relaxation time $\tau_{sf}$ are determined by the relation $C_0=N_FD_0$ and $\lambda=\sqrt{2\tau_{sf}D_0(1-\beta\beta')}$, where 
we set the density of states at the Fermi level as $N_F=7.5\text{nm}^{-3}\text{eV}^{-1}$.

\begin{figure}[p]
\includegraphics[width=0.8\columnwidth]{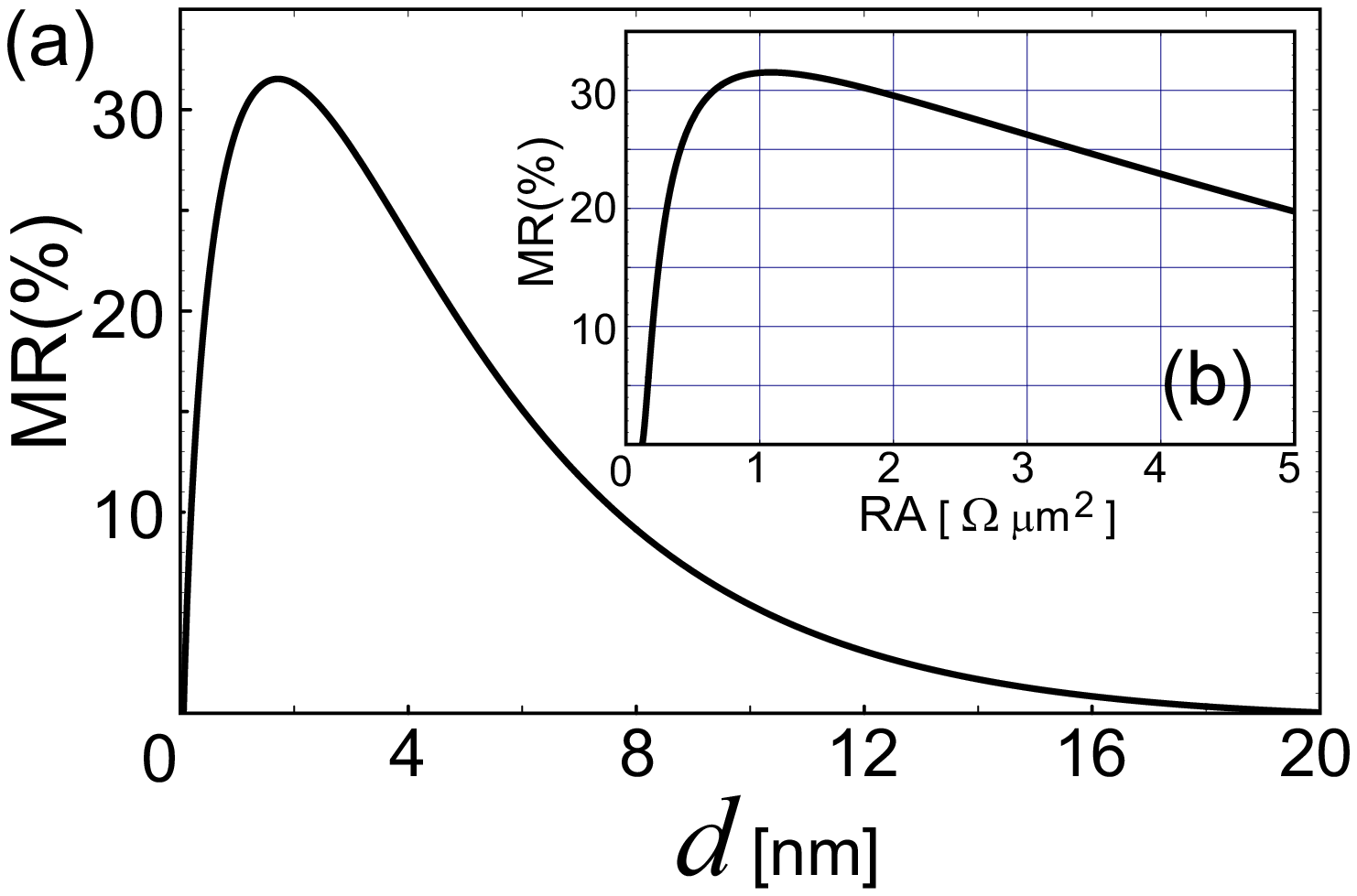}
\caption{
  MR ratios of the domain wall CCP spin-valve are plotted 
  with varying the contact diameter $d$. 
  (a) MR is plotted against contact diameter $d$. 
  (b) MR is plotted against resistance area product RA. 
      The value of RA without CCP structure is set at $0.1\Omega\mu\text{m}^2$. 
}
\label{fig2}
\end{figure}

In order to examine the MR enhancement effect by the CCP structure, 
we first vary the contact diameter $d$ with the fixed size of the electrode $D=40\text{nm}$. 
The dependence of MR on $d$ is shown in Fig. \ref{fig2} (a). 
The MR enhancement effect by the CCP structure is confirmed by the simulation 
as is shown in the figure, 
where the MR ratio increases with decreasing the contact diameter 
$d$ from $d=20$nm to $d=2$nm. 
If we completely confine the contact region $d\to0$, no electric current can flow and 
MR ratio becomes zero. This implies the existence of the contact diameter 
which maximize the MR ratio. 
Actually we observe sharp peak around $d\sim$2 nm in Fig. \ref{fig2} (a). 
In practice, in the region where the contact diameter is less than the order of 1 nm, 
the diffusion equation loses its validity and 
the quantum mechanical approach is necessary. 
In Fig. \ref{fig2} (b) we replot the MR ratio against the 
resistance area product RA. 
The value of RA without CCP structure is set at $0.1\Omega\mu\text{m}^2$. 
It is observed that 
if we decrease the contact diameter, the MR increases with increasing RA 
and takes its maximum value around RA$\sim1\Omega\mu\text{m}^2$. 

\begin{figure}[p]
\includegraphics[width=0.8\columnwidth]{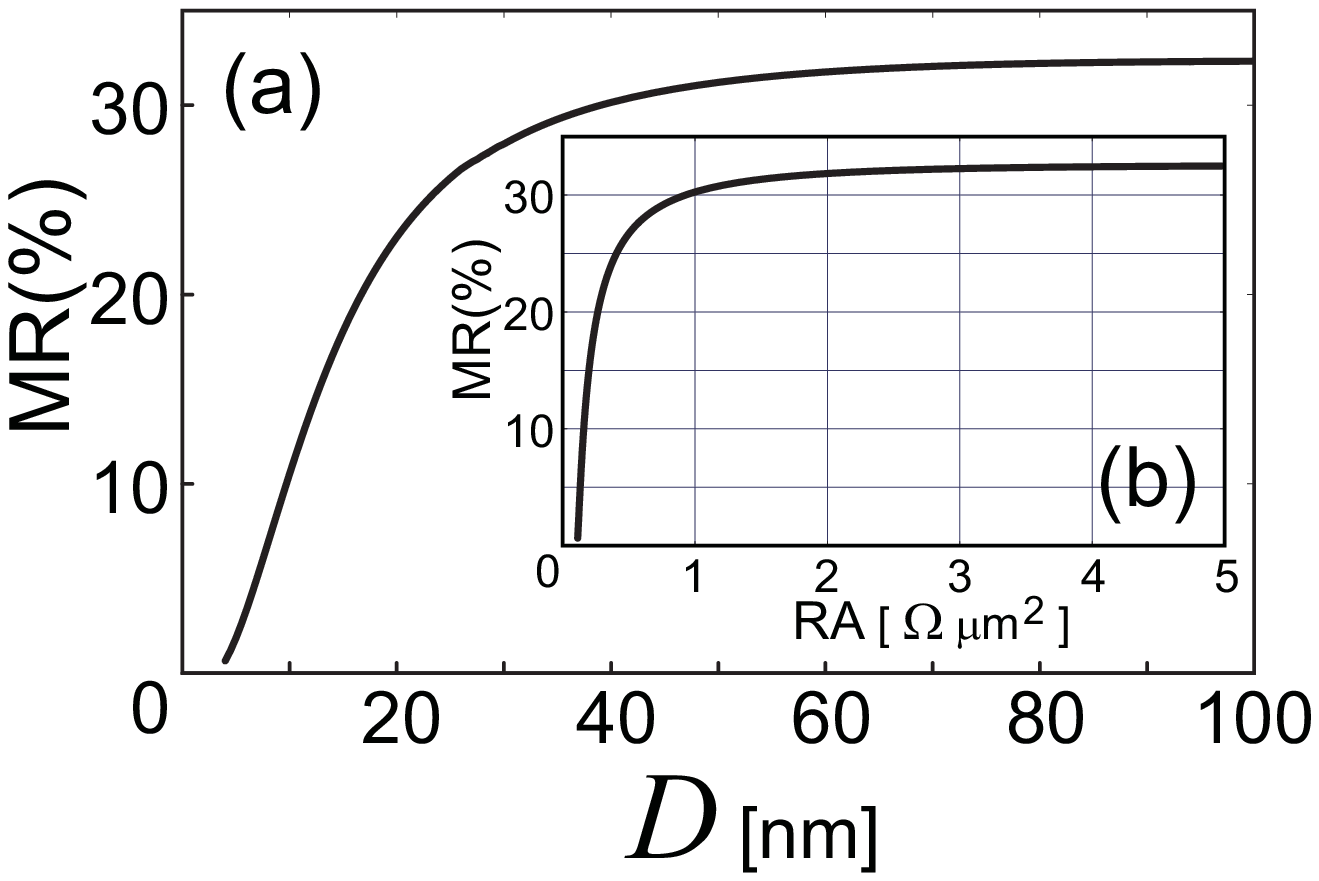}
\caption{
  MR ratios of the domain wall CCP spin-valve are plotted 
  with varying the size of the electrode $D$. 
  (a) MR is plotted against the size of the electrode $D$. 
  (b) MR is plotted against resistance area product RA. 
      The value of RA without CCP structure is set at $0.1\Omega\mu\text{m}^2$. 
}
\label{fig3}
\end{figure}

Next we vary the size of the electrode $D$ with the fixed contact diameter $d$=2nm. 
Since enlarging the size of the electrode $D$ corresponds to 
increasing the distance between nanocontacts in the NOL, 
the effect of changing the density of nanocontacts 
can be realized in our simulation by varying the size of the electrode. 
The dependence of MR on $D$ is shown in Fig. \ref{fig3} (a). 
The MR is a monotonic increase function of $D$ 
and is saturated around $D\sim$50nm as is shown in the figure. 
This means that the MR monotonically increases with decreasing the density of nanocontacts 
 and is saturated at a certain value of density. 
In Fig. \ref{fig3} (b) we replot the MR ratio against the 
resistance area product RA. 
Since RA increases with decreasing the density of nanocontacts, 
we observe a monotonic increasing dependence of MR on RA 
unlike the case with varying the contact radius $d$ in Fig. \ref{fig2} (b) 
where MR has maximum value at RA$\sim1\Omega\mu\text{m}^2$. 
Although the MR shows monotonic increasment with RA, 
it is saturated around RA$=1\sim2\,\Omega\mu\text{m}^2$. 

In summary, we studied the current-perpendicular-to-plane magnetoresistance 
of a domain wall confined in a CCP structure 
made of an NOL. 
We solved the continuity equations for charge and spin currents 
with the three-dimensional CCP geometry by use of finite
element method. 
We have confirmed that the MR enhancement effect by the CCP structure 
from the calculation of MR with varying the contact radius $d$ and 
the electrode size $D$. 
We observed that the MR is maximized at $d\sim$2nm. 
In our simulation with varying the electrode size $D$, 
it is observed that the MR increases with decreasing the density of nanocontact and 
is saturated at a certain value of density.

%
\section*{Acknowledgement}
The authors thank M.~Sahashi, M.~Doi, H.~Iwasaki, M.~Takagishi,
Y.~Rikitake and K.~Seki for valuable discussions.  The work
has been supported by The New Energy and Industrial Technology
Development Organization (NEDO).

\end{document}